\documentclass[aps,prl,twocolumn,superscriptaddress,groupedaddress,showpacs,showkeys,floatfix]{revtex4-1}

\usepackage{graphicx}
\usepackage{hyperref}
\usepackage[caption=false]{subfig}
\usepackage{soul}
\usepackage{color}

\begin{document}

\newcommand{\leftexp}[2]{{\vphantom{#2}}^{#1}{#2}}
\newcommand{\leftboth}[3]{{\vphantom{#3}}^{#1}_{#2}{#3}}

\title{Measurement of coherent $\pi^{+}$ production in low energy neutrino-Carbon scattering}

% Authors

\newcommand{\INSTEE}{\affiliation{University of Bern, Albert Einstein Center for Fundamental Physics, Laboratory for High Energy Physics (LHEP), Bern, Switzerland}}
\newcommand{\INSTFE}{\affiliation{Boston University, Department of Physics, Boston, Massachusetts, U.S.A.}}
\newcommand{\INSTD}{\affiliation{University of British Columbia, Department of Physics and Astronomy, Vancouver, British Columbia, Canada}}
\newcommand{\INSTGA}{\affiliation{University of California, Irvine, Department of Physics and Astronomy, Irvine, California, U.S.A.}}
\newcommand{\INSTI}{\affiliation{IRFU, CEA Saclay, Gif-sur-Yvette, France}}
\newcommand{\INSTGB}{\affiliation{University of Colorado at Boulder, Department of Physics, Boulder, Colorado, U.S.A.}}
\newcommand{\INSTFG}{\affiliation{Colorado State University, Department of Physics, Fort Collins, Colorado, U.S.A.}}
\newcommand{\INSTFH}{\affiliation{Duke University, Department of Physics, Durham, North Carolina, U.S.A.}}
\newcommand{\INSTBA}{\affiliation{Ecole Polytechnique, IN2P3-CNRS, Laboratoire Leprince-Ringuet, Palaiseau, France }}
\newcommand{\INSTEF}{\affiliation{ETH Zurich, Institute for Particle Physics, Zurich, Switzerland}}
\newcommand{\INSTEG}{\affiliation{University of Geneva, Section de Physique, DPNC, Geneva, Switzerland}}
\newcommand{\INSTDG}{\affiliation{H. Niewodniczanski Institute of Nuclear Physics PAN, Cracow, Poland}}
\newcommand{\INSTCB}{\affiliation{High Energy Accelerator Research Organization (KEK), Tsukuba, Ibaraki, Japan}}
\newcommand{\INSTED}{\affiliation{Institut de Fisica d'Altes Energies (IFAE), The Barcelona Institute of Science and Technology, Campus UAB, Bellaterra (Barcelona) Spain}}
\newcommand{\INSTEC}{\affiliation{IFIC (CSIC \& University of Valencia), Valencia, Spain}}
\newcommand{\INSTEI}{\affiliation{Imperial College London, Department of Physics, London, United Kingdom}}
\newcommand{\INSTGF}{\affiliation{INFN Sezione di Bari and Universit\`a e Politecnico di Bari, Dipartimento Interuniversitario di Fisica, Bari, Italy}}
\newcommand{\INSTBE}{\affiliation{INFN Sezione di Napoli and Universit\`a di Napoli, Dipartimento di Fisica, Napoli, Italy}}
\newcommand{\INSTBF}{\affiliation{INFN Sezione di Padova and Universit\`a di Padova, Dipartimento di Fisica, Padova, Italy}}
\newcommand{\INSTBD}{\affiliation{INFN Sezione di Roma and Universit\`a di Roma ``La Sapienza'', Roma, Italy}}
\newcommand{\INSTEB}{\affiliation{Institute for Nuclear Research of the Russian Academy of Sciences, Moscow, Russia}}
\newcommand{\INSTHA}{\affiliation{Kavli Institute for the Physics and Mathematics of the Universe (WPI), The University of Tokyo Institutes for Advanced Study, University of Tokyo, Kashiwa, Chiba, Japan}}
\newcommand{\INSTCC}{\affiliation{Kobe University, Kobe, Japan}}
\newcommand{\INSTCD}{\affiliation{Kyoto University, Department of Physics, Kyoto, Japan}}
\newcommand{\INSTEJ}{\affiliation{Lancaster University, Physics Department, Lancaster, United Kingdom}}
\newcommand{\INSTFC}{\affiliation{University of Liverpool, Department of Physics, Liverpool, United Kingdom}}
\newcommand{\INSTFI}{\affiliation{Louisiana State University, Department of Physics and Astronomy, Baton Rouge, Louisiana, U.S.A.}}
\newcommand{\INSTJ}{\affiliation{Universit\'e de Lyon, Universit\'e Claude Bernard Lyon 1, IPN Lyon (IN2P3), Villeurbanne, France}}
\newcommand{\INSTHB}{\affiliation{Michigan State University, Department of Physics and Astronomy,  East Lansing, Michigan, U.S.A.}}
\newcommand{\INSTCE}{\affiliation{Miyagi University of Education, Department of Physics, Sendai, Japan}}
\newcommand{\INSTDF}{\affiliation{National Centre for Nuclear Research, Warsaw, Poland}}
\newcommand{\INSTFJ}{\affiliation{State University of New York at Stony Brook, Department of Physics and Astronomy, Stony Brook, New York, U.S.A.}}
\newcommand{\INSTGJ}{\affiliation{Okayama University, Department of Physics, Okayama, Japan}}
\newcommand{\INSTCF}{\affiliation{Osaka City University, Department of Physics, Osaka, Japan}}
\newcommand{\INSTGG}{\affiliation{Oxford University, Department of Physics, Oxford, United Kingdom}}
\newcommand{\INSTBB}{\affiliation{UPMC, Universit\'e Paris Diderot, CNRS/IN2P3, Laboratoire de Physique Nucl\'eaire et de Hautes Energies (LPNHE), Paris, France}}
\newcommand{\INSTGC}{\affiliation{University of Pittsburgh, Department of Physics and Astronomy, Pittsburgh, Pennsylvania, U.S.A.}}
\newcommand{\INSTFA}{\affiliation{Queen Mary University of London, School of Physics and Astronomy, London, United Kingdom}}
\newcommand{\INSTE}{\affiliation{University of Regina, Department of Physics, Regina, Saskatchewan, Canada}}
\newcommand{\INSTGD}{\affiliation{University of Rochester, Department of Physics and Astronomy, Rochester, New York, U.S.A.}}
\newcommand{\INSTHC}{\affiliation{Royal Holloway University of London, Department of Physics, Egham, Surrey, United Kingdom}}
\newcommand{\INSTBC}{\affiliation{RWTH Aachen University, III. Physikalisches Institut, Aachen, Germany}}
\newcommand{\INSTFB}{\affiliation{University of Sheffield, Department of Physics and Astronomy, Sheffield, United Kingdom}}
\newcommand{\INSTDI}{\affiliation{University of Silesia, Institute of Physics, Katowice, Poland}}
\newcommand{\INSTEH}{\affiliation{STFC, Rutherford Appleton Laboratory, Harwell Oxford,  and  Daresbury Laboratory, Warrington, United Kingdom}}
\newcommand{\INSTCH}{\affiliation{University of Tokyo, Department of Physics, Tokyo, Japan}}
\newcommand{\INSTBJ}{\affiliation{University of Tokyo, Institute for Cosmic Ray Research, Kamioka Observatory, Kamioka, Japan}}
\newcommand{\INSTCG}{\affiliation{University of Tokyo, Institute for Cosmic Ray Research, Research Center for Cosmic Neutrinos, Kashiwa, Japan}}
\newcommand{\INSTGI}{\affiliation{Tokyo Metropolitan University, Department of Physics, Tokyo, Japan}}
\newcommand{\INSTF}{\affiliation{University of Toronto, Department of Physics, Toronto, Ontario, Canada}}
\newcommand{\INSTB}{\affiliation{TRIUMF, Vancouver, British Columbia, Canada}}
\newcommand{\INSTG}{\affiliation{University of Victoria, Department of Physics and Astronomy, Victoria, British Columbia, Canada}}
\newcommand{\INSTDJ}{\affiliation{University of Warsaw, Faculty of Physics, Warsaw, Poland}}
\newcommand{\INSTDH}{\affiliation{Warsaw University of Technology, Institute of Radioelectronics, Warsaw, Poland}}
\newcommand{\INSTFD}{\affiliation{University of Warwick, Department of Physics, Coventry, United Kingdom}}
\newcommand{\INSTGE}{\affiliation{University of Washington, Department of Physics, Seattle, Washington, U.S.A.}}
\newcommand{\INSTGH}{\affiliation{University of Winnipeg, Department of Physics, Winnipeg, Manitoba, Canada}}
\newcommand{\INSTEA}{\affiliation{Wroclaw University, Faculty of Physics and Astronomy, Wroclaw, Poland}}
\newcommand{\INSTH}{\affiliation{York University, Department of Physics and Astronomy, Toronto, Ontario, Canada}}

\INSTEE
\INSTFE
\INSTD
\INSTGA
\INSTI
\INSTGB
\INSTFG
\INSTFH
\INSTBA
\INSTEF
\INSTEG
\INSTDG
\INSTCB
\INSTED
\INSTEC
\INSTEI
\INSTGF
\INSTBE
\INSTBF
\INSTBD
\INSTEB
\INSTHA
\INSTCC
\INSTCD
\INSTEJ
\INSTFC
\INSTFI
\INSTJ
\INSTHB
\INSTCE
\INSTDF
\INSTFJ
\INSTGJ
\INSTCF
\INSTGG
\INSTBB
\INSTGC
\INSTFA
\INSTE
\INSTGD
\INSTHC
\INSTBC
\INSTFB
\INSTDI
\INSTEH
\INSTCH
\INSTBJ
\INSTCG
\INSTGI
\INSTF
\INSTB
\INSTG
\INSTDJ
\INSTDH
\INSTFD
\INSTGE
\INSTGH
\INSTEA
\INSTH

\author{K.\,Abe}\INSTBJ
\author{C.\,Andreopoulos}\INSTEH\INSTFC
\author{M.\,Antonova}\INSTEB
\author{S.\,Aoki}\INSTCC
\author{A.\,Ariga}\INSTEE
\author{S.\,Assylbekov}\INSTFG
\author{D.\,Autiero}\INSTJ
\author{S.\,Ban}\INSTCD
\author{M.\,Barbi}\INSTE
\author{G.J.\,Barker}\INSTFD
% ...

\collaboration{The T2K Collaboration}\noaffiliation

\date{\today}

\begin{abstract}
We report the first measurement of the flux-averaged cross section for charged current coherent
$\pi^{+}$ production on carbon for neutrino energies less than 1.5~GeV .
%\st{to a restricted final state phase space region}
and with a restriction on the final state phase space volume
in the T2K near detector, ND280. 
Comparisons are made with predictions from the Rein-Sehgal coherent production model and the
model by Alvarez-Ruso {\it et al.}, the latter representing the first implementation of an instance of the new class 
of microscopic coherent models in a neutrino interaction Monte Carlo event generator.
We observe a clear event excess above background, disagreeing with the
null results reported by K2K and SciBooNE in a similar neutrino energy region. The measured flux-averaged cross sections 
are below those predicted by both the Rein-Sehgal and the Alvarez-Ruso {\it et al.} models.
\end{abstract}


\maketitle

{\it Introduction}\textemdash Charged current coherent pion production in neutrino-nucleus 
scattering, $\nu_{\mu} + A \rightarrow \mu^{-} + \pi^{+} + A$,
is a process in which the neutrino scatters coherently from an entire nucleus, leaving the nucleus unchanged.
%Preservation of nuclear coherence requires that no nucleon be singled out in the interaction. Thus,
No quantum numbers are exchanged and there is little four-momentum 
transfer to any nucleon. 
Due to these restrictions the outgoing lepton and pion are aligned with the beam direction
and no other hadrons are produced.

\begin{figure}[t]
\subfloat[]
{
\includegraphics[scale=0.35]{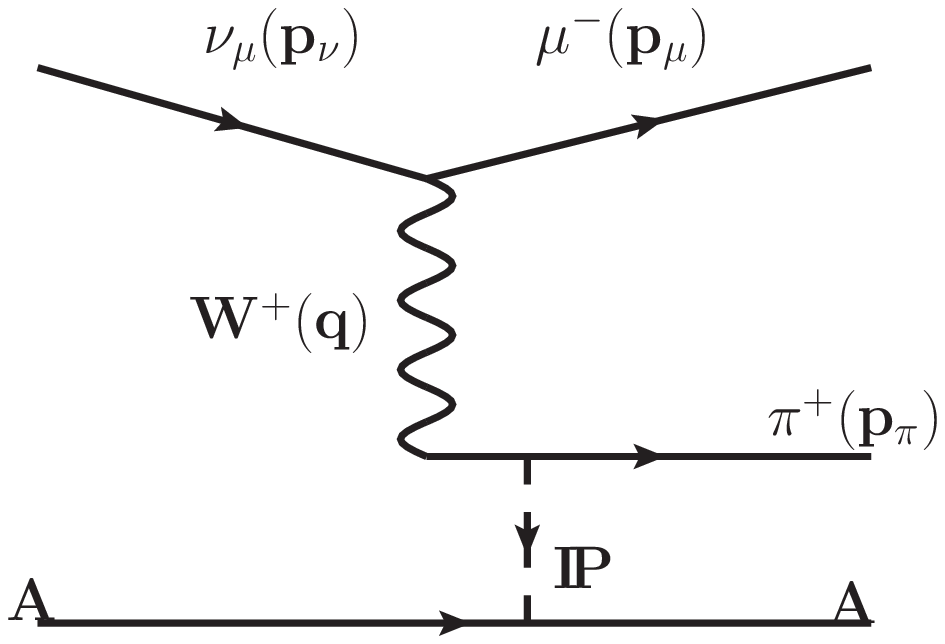}
\label{Theory:RS}
}
\qquad
\subfloat[]
{
\includegraphics[scale=0.35]{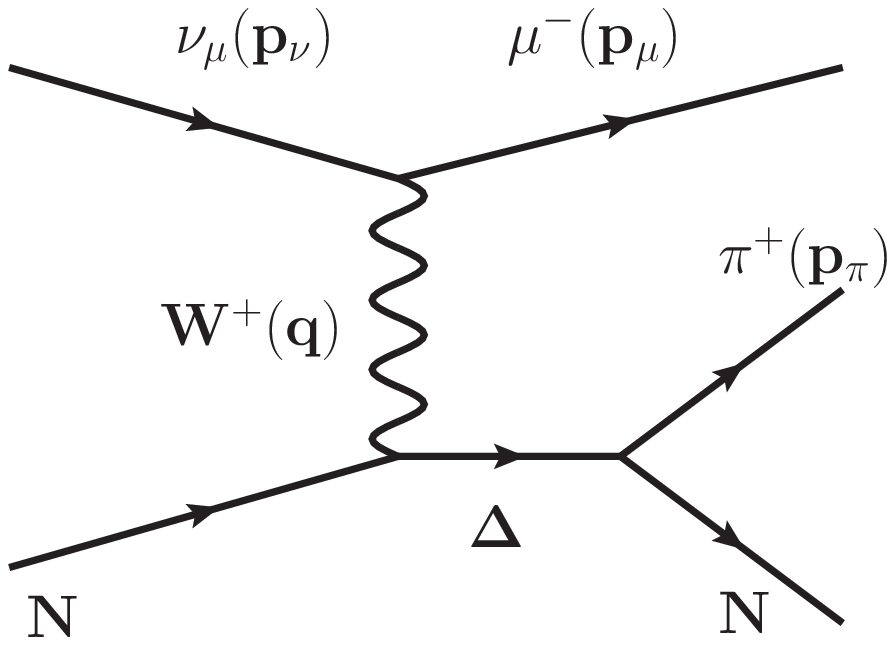}
\label{Theory:AR}
}
\caption{\label{fig:Theory} (a) the diagram for coherent charged
pion production model based on Adler's Theorem. The {I$\!$P} represents the transfer of a Pomeron to
the nuclear system. (b) Dominant diagram for the microscopic class of
coherent charged pion production models.}
\end{figure}

Two classes of models have been developed to describe this process. 
The first class uses Adler's theorem~\cite{Adler} to relate the coherent scattering
cross section at $Q^{2} = -q^{2} = -(p_{\nu} - p_{\mu})^{2} = 0$ with the pion-nucleus 
elastic scattering cross section. 
Described by the diagram shown in Fig.~\ref{fig:Theory}\subref{Theory:RS}, the differential cross-section is
\begin{equation}
\frac{d\sigma_{coh}}{dQ^{2} dy d|t|}\biggr \vert_{Q^{2}=0} = \frac{G_{F}^{2}}{2 \pi^{2}}  f_{\pi}^{2} \frac{1-y}{y} \frac{d\sigma( \pi A \rightarrow \pi A)}{d|t|},
\end{equation}
where $y = E_{\pi} / E_{\nu}$ with $E_{\pi}$ and $E_{\nu}$ being the energy of the pion and neutrino respectively, 
$f_{\pi}$ is the pion decay constant and $|t| = | (q - p_{\pi}) |^{2}$ is the magnitude of the square of the 
four-momentum transferred by the exchange boson
to the nucleus. Different models~\cite{Kopeliovich1, BergerSehgal, PaschosSchalla, Andreopoulos:2009rq} 
choose different methods for extension to $Q^{2} > 0$ and implementations of the $\pi A$ elastic scattering
cross-section.
The validity of these models below neutrino energies of roughly 
2\,GeV is limited~\cite{ReinSehgalCoh,Hernandez:2009vm,Hernandez:2010jf,Amaro:2008hd}.

The second class, known as the {\em microscopic} models, was developed specifically for neutrino energies 
less than 2\,GeV~\cite{Singh:2006bm,AlvarezRuso:2007tt,Leitner:2009ph,Nakamura:2009iq,Hernandez:2010jf}.
These models are based on the single nucleon process $\nu_{l} N \rightarrow l^{-} N \pi^{+}$, which
is dominated by $\Delta$ production at low energies as shown in the right diagram in Fig.~\ref{fig:Theory}\subref{Theory:AR}. 
The total cross
section is derived from the coherent sum of the contribution of all nucleons within the individual nuclei. Effects
of the nuclear medium on the $\Delta$ and on the pion wavefunction are taken into account. These models
have not been tested against data. Only recently has one instance of this class,
the model from Alvarez-Ruso {\it et al.}~\cite{AlvarezRuso:2007tt}, been implemented in a neutrino event generator.
%\end{itemize}

The charged current coherent production cross section has been measured at neutrino energies above 7\,GeV by several
experiments~\cite{Allport:1988cq,Vilain:1993sf,Grabosch:1985mt,Willocq:1992fv,Acciarri:2014eit} and has been found to agree with 
the standard coherent model developed by Rein and Sehgal. 
More recent model dependent searches by K2K~\cite{Hasegawa:2005td} and SciBooNE~\cite{Hiraide:2008eu} at neutrino energies of
0.5 - 2\,GeV suffer from low statistics and reported null results. 
%signals that were consistent with large backgrounds at the 90\% confidence level.
Recently the MINER$\nu$A experiment published a measurement of this cross section for neutrino energies between
1.5\,GeV and 20.0\,GeV~\cite{Higuera:2014azj}. 

This letter presents the first measurement of the charged current coherent pion production cross section below
a neutrino energy of 1.5\,GeV. The analysis conducts a model independent search for an excess of events 
at low $|t|$.
The flux averaged charged current coherent pion production cross section is presented for two regions of
the final state phase space. The restricted final state phase space region is limited to 
$p_{\mu, \pi} > 0.18\,\mbox{GeV/c}$, 
$\theta_{\mu,\pi} < 70^{\circ}$, which removes areas of low detector acceptance, and $p_{\pi} < 1.6\,\mbox{GeV/c}$, which removes
an area outside the range of validity of the microscopic model. The angles of the muon
and pion, $\theta_{\mu,\pi}$, are measured with respect to the average direction of the incoming neutrino beam.

The flux averaged cross section for production to the complete phase space is also
presented. In addition, for each choice of final state phase space coverage, we present results using 
two different models: the Rein-Sehgal model~\cite{ReinSehgalCoh} as implemented in the GENIE 2.6.4 neutrino event generator (which uses
a more sophisticated parameterisation of the pion-nucleus elastic scattering than outlined in the original Rein-Seghal
paper\footnote{ Note that the official version of GENIE 2.6.4
incorporates an error in the calculation of the pion-nucleus cross section. This error was fixed in GENIE versions
2.8.2 and beyond. The error has also been fixed for the version of 2.6.4 used in the current analysis.})
and an implementation of the microscopic model constructed by Alvarez-Ruso {\it et al.}~\cite{AlvarezRuso:2007tt}. 
Previous null results~\cite{Hasegawa:2005td, Hiraide:2008eu} used the Rein-Seghal coherent model
to devise and tune kinematic cuts and were, thus, not model independent. 

{\it T2K Experiment}\textemdash T2K~\cite{Abe:2011ks} is an off-axis long-baseline neutrino oscillation experiment
sited at the J-PARC facility in Tokai, Japan. A complete description of the experiment may be found 
in Ref.~\cite{Abe:2011ks}. The experiment views an off-axis neutrino beam flux that is composed of 92.6\% $\nu_{\mu}$
with a peak $\nu_{\mu}$ energy of 0.6\,GeV. Details of the neutrino beam are described in detail in references~\cite{Abe:2011ks} and \cite{Abe:2012av}.  The data used in this analysis corresponds to $5.54 \times 10^{20}$ protons
on target (POT).

ND280~\cite{Abe:2011ks} is  the off-axis magnetized tracking near detector designed 
to measure interactions 
of both $\nu_{\mu}$ and $\nu_{e}$ from
the T2K beam before oscillations. The detector rests within the refurbished UA1/NOMAD magnet, which provides a magnetic
field of 0.2~T, and is split into two regions: the upstream $\pi^{0}$ detector~\cite{Assylbekov:2011sh} 
and the tracker. The tracker
region contains two plastic scintillator detectors~\cite{Amaudruz:2012esa} (FGDs or Fine Grained Detectors), used 
as targets for neutrino
interactions, sandwiched between three argon-gas TPCs~\cite{Abgrall:2010hi}. The first, most upstream, FGD (FGD1),
only has layers of plastic ($CH$) scintillator bars whilst the second FGD (FGD2) also contains water layers. 
Surrounding these inner
subdetectors is a set of electromagnetic calorimeters~\cite{Allan:2013ofa}.
%which  increase  the
%hermeticity of the detector and tag outgoing particles.
The magnet yokes are instrumented with 
scintillator-based side muon range detectors~\cite{Aoki:2012mf} to track high angle muons.

Neutrino interactions are simulated using the default GENIE 2.6.4 neutrino event generator 
package~\cite{Andreopoulos:2009rq}. 
Quasielastic scattering is modelled using the Llewellyn-Smith~\cite{LlewllynSmith} model with an
axial mass, $m_{A}$, set to $0.99\,\mbox{GeV/c}^{2}$.
The initial state nuclear model is the Bodek-Ritchie relativistic Fermi gas model (RFG) with a 
Fermi momentum of 221\,MeV/c, extended 
to include short range nucleon-nucleon correlations~\cite{Bodek:1980ar}. Inelastic single
pion production from resonances is simulated using the Rein-Sehgal model~\cite{Rein:1980wg}.
Interference between the resonance states and lepton mass effects are ignored,
although the effect 
of lepton masses on phase space boundaries is taken into account. Non-resonant pion production is modelled using an
extension of the Bodek-Yang model~\cite{Bodek:2004pc} to low energies. Interference between the resonant and
non-resonant interaction modes is not taken into account. 
The relative contributions were tuned by GENIE 
against available single pion production cross section data~\cite{Andreopoulos:2009rq}.
The transition to non-resonant inelastic
scattering is simulated using the same Bodek-Yang model. 
Hadronisation is described using the AGKY model~\cite{Yang:2009zx}.
Hadronic interactions in the nuclear medium are modelled 
using the INTRANUKE package~\cite{Andreopoulos:2009rq}.

{\it Event selection}\textemdash This analysis uses neutrino interactions which have occurred in the scintillator
target of FGD1. Charged particles in the final state are analysed by the second TPC, which lies immediately downstream
of FGD1. 
The first step is to select $\nu_{\mu}$ CC inclusive events in FGD1 using the event selection criteria
reported in detail in Ref.~\cite{Abe:2013jth}. Events passing this selection are in-time with the beam and 
contain at least one negatively charged track in TPC2 consistent with a minimally ionising particle. 
The interaction vertex is defined to be the most upstream point of the muon candidate track. This must lie within
the fiducial volume of FGD1, which excludes the two most upstream and downstream layers,
and the outer-most 5 bars in each layer. 
All previously published results do not use vertex activity and do not impose such a constraint on the downstream fiducial boundary.
The resulting fiducial region contains 0.74
tonnes of carbon\footnote{FGD1 also contains hydrogen. Pure diffractive scattering from the protons can
occur which, at low $Q^{2}$, may result in a similar final state to that produced by coherent interactions on nuclei. 
The contribution of diffractive scattering from hydrogen to the selected event sample
was estimated to be less than 5\%.}.

An event sample with an enhanced coherent pion component is then selected by requiring
a second, positively charged, track originating from the interaction vertex. 
This second track is required to have a $dE/dx$ profile
consistent with a muon or pion traversing the TPC. Cuts to enforce this requirement
remove proton tracks such that they make up less than 3\% of the selected pion candidates.

Charged current coherent pion production leaves the nuclear target unchanged and in its ground state. 
Hence the only particles exiting the interaction are
a charged lepton and an oppositely charged pion. 
Events with additional energy deposited around the vertex are removed by a cut on the
vertex activity (VA), which is defined to be the sum of all energy deposits within a cubic 
volume with side length 5~cm centered on the vertex. 
No attempt is made to estimate and
subtract the energy deposited by the muon and pion within this region.
Simulated coherent events typically deposit 220\,PEU (Photon Equivalent Unit\footnote{
A Photon Equivalent Unit is a measure of the response of the FGD to single photons. A single PEU corresponds
to a deposited energy of 0.046\,MeV.})
with an RMS spread of 40\,PEU. 
Sixty percent of the predicted
background is removed by requiring the VA in the event to be less than 300\,PEU with no loss of predicted signal.

{\it Analysis strategy}\textemdash In the models based on Adler's theorem, coherent interactions are 
characterised by the low transfer of four-momentum to the nucleus. Referring to
the diagram in Fig.~\ref{fig:Theory}\subref{Theory:RS}, this quantity is defined to be
\begin{equation}
|t| = | (q - p_{\pi})^{2} | = \left( \sum_{i=\mu,\pi} (E_{i} - p_{i}^{L}) \right)^{2} + \left( \sum_{i=\mu,\pi} p_{i}^{T} \right)^{2}
\end{equation}
where the approximation that negligible energy is transferred to the nucleus has been made, and $p^{T}$ 
and $p^{L}$ are the transverse
and longitudinal components of the particle's momentum with respect to the direction of the neutrino beam. 
The microscopic models also predict events clustering at low values of $|t|$. This experimental observable
is well defined regardless of the class of model under consideration.
This analysis searches for an excess of events above background at low $|t|$.
No attempt is made to fit any particular model to the data.

{\it Sources of systematic uncertainty}\textemdash The flux averaged cross section is given by 
$\langle \sigma_{coh} \rangle = ( N_{sel} - N_{bg} ) / \Phi N_{T} \epsilon$ where $N_{sel}$ is the number of selected events, $N_{bg}$ is the
estimated number of background events, $\epsilon$ is the coherent event selection
efficiency, $N_{T}$ is the number of target carbon nuclei and $\Phi$ is the integrated T2K neutrino flux incident on FGD1.
The largest uncertainties on the flux-averaged cross section arise from: the flux model,
the background interaction model, the model for final state pion reinteractions within the detector, 
and the model for the VA. Estimates of the uncertainty on the coherent cross section 
are determined by varying model parameters within their uncertainties, and propagating the changes to the result.

The flux systematic uncertainty is evaluated by varying the shape 
and normalisation of the T2K flux prediction~\cite{Abe:2012av}.
The uncertainties in the parameters of the background cross section models are constrained by 
previous measurements as implemented in the default configuration of 
the GENIE generator~\cite{Andreopoulos:2009rq,Rodrigues:2016xjj}.
The pion reinteraction uncertainty is evaluated by varying the total pion absorption and 
charge exchange cross sections within
bounds defined by the difference between GEANT4 and published hadronic interaction data~\cite{Ashery:1981tq}.

The VA uncertainty arises from two sources: the charge response of the FGD to energy deposition
and the simulation of energy produced at the vertex in the charged current coherent $\pi^{+}$ background event sample.
The former was studied by comparing the charge response of the FGD to protons stopping in the FGD 
fiducial volume in data and Monte Carlo. The simulation was found to underestimate the average measured charge deposit
by 10\% and this was taken to be the systematic uncertainty in the FGD charge response.

The average VA of the simulated coherent background control sample was lower than that observed in the data.
The issue of multi-nucleon knockout effects in neutrino scattering has recently received much attention (see, 
for example,~\cite{Martini:2010ex,Nieves:2005rq}). Such effects would eject low momentum protons into the region 
around the vertex, increasing the average VA. Indeed, the simulated VA distribution can be made 
to agree better with background data by adding VA consistent with that deposited by a proton with kinetic energy distributed uniformly
between 20 and 225\,MeV to 25\% of background events with a neutron target.
The MINER$\nu$A experiment reported a similar observation in a study of neutrino-nucleus
quasi-elastic interactions~\cite{Fiorentini:2013ezn,Rodrigues:2015hik}.
The uncertainty in the vertex activity model was derived by switching 
this addition on and off. No correction is applied for this effect in deriving the cross section or significance
of the signal. This is the dominant systematic uncertainty in the estimate of the background 
to the charged current coherent $\pi^{+}$ signal.

{\it Background estimate}\textemdash The estimated number of background interactions is constrained 
by fits to the data. The event sample was divided into a signal enriched sample, 
with $|t| < 0.15\,\mbox{(GeV/c)}^{2}$ and $\mbox{VA} < 300\,\mbox{PEU}$; and two side-band regions.
The first side-band is comprised of events which fail the VA cut 
($|t| < 0.15\,\mbox{(GeV/c)}^{2}$ and $\mbox{VA} > 300\,\mbox{PEU}$), while the second region 
contains events which fail the $|t|$ cut ($|t| > 0.15\,\mbox{(GeV/c)}^{2}$ and $\mbox{VA} < 300\,\mbox{PEU}$).
The Monte Carlo predicts a $|t|$ resolution for signal events of less than $0.02~(\mbox{GeV/c})^{2}$. 
The signal enriched region was defined to include more than 99\% of the coherent signal predicted by either model.
Events in the side-band samples were then sorted into bins of reconstructed invariant mass, $W$. 
Template distributions of pion momenta were formed for each $W$ bin and scale factors
estimated by fitting the normalisation of each $W$ bin to the data. The variation in $W$ was constrained by the 
covariance matrices encoding the effects of the variation in the systematic parameters described above.

The scale factors resulting from the fit to the side-bands were constant at 89\% over the full $W$-range. 
The pre-fit incoherent background prediction was thereby reduced from 88 events to $78 \pm 18$
The fractional uncertainties in the background estimate from these sources of uncertainty are shown in 
Table \ref{tab:Exp:syst}.

\begin{table}
\centering
\begin{tabular}{ccc} \hline \hline
Systematic Source     &  Fractional error & Fractional error  \\ 
                      &  on background   &  on $\langle \sigma_{coh}^{rest} \rangle$  \\ \hline
Flux model            & 0.05  &   0.10         \\
Background model      & 0.14   &  0.25         \\
Pion reinteractions   & +0.05~-0.01   &  +0.14~-0.05   \\
Vertex activity model & 0.19   &   0.28         \\ 
FGD Charge scale      & 0.06   &  0.15          \\  \hline \hline
\end{tabular}
\caption{ Summary of the fractional systematic uncertainties on the background estimate and on the phase space restricted charged
current coherent flux averaged cross section ( $\langle \sigma_{coh}^{rest} \rangle$).}
\label{tab:Exp:syst}
\end{table}

\begin{figure}[t]
\includegraphics[width=0.8\columnwidth]{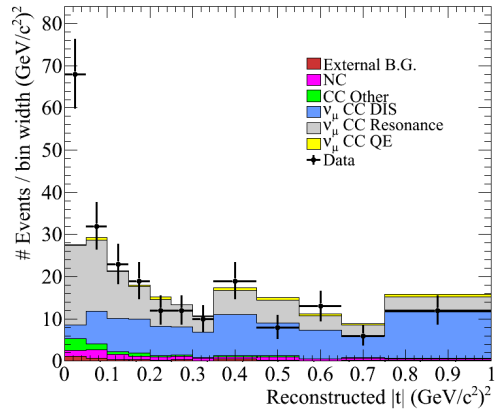}
\caption{\label{fig:Exp:t} The reconstructed $|t|$ distribution after the VA cut and the background
tuning procedure have been applied. The errors on the data are statistical only, and the uncertainty on the
tuned model is not shown.
The model's prediction of the coherent contribution has been removed from the plot. The small external
background component contains events that occur outside the FGD1 fiducial volume, such as
interactions occurring in the surrounding magnet volume.}
\end{figure}

{\it Results}\textemdash 
The distribution of $|t|$ for the data and the predicted background, both after the VA
cut is applied, is shown in Fig.~\ref{fig:Exp:t}.
There is a clear excess of events in the data at low $|t|$ that is consistent
with a charged current coherent $\pi^{+}$ production signal, while the shape of the high $|t|$ region is 
consistent with the background prediction.
The total number of events observed in
the signal region in the data is 123. After background subtraction, the
number of coherent events in the data is $45 \pm 18$. The significance of observing such an excess of 
events is $2.2\,\sigma$ with a p-value of 0.014.

\begin{figure}[t]
\includegraphics[width=0.8\columnwidth]{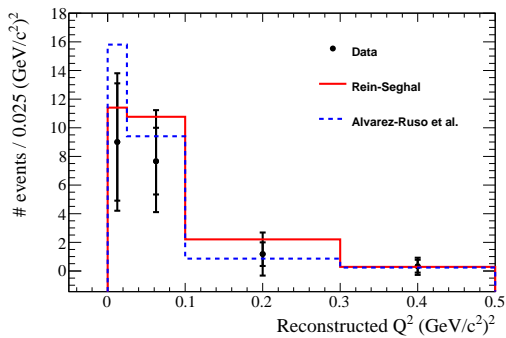}
\caption{\label{fig:Exp:Q2} The reconstructed $Q^{2}$ distribution after background subtraction. The inner error
bars represent the statistical uncertainty on the data before background subtraction and the outer the total
uncertainty which also includes systematic effects. Correlations between bins are not reflected in the uncertainty displayed
on the figure. The last bin is an overflow bin, containing
all events with reconstructed $Q^{2}$ greater than $0.3\,(\mbox{GeV/c})^{2}$.}
\end{figure}

The model-dependent efficiency for selecting coherent events in the restricted phase space ($p_{\mu, \pi} > 0.18\,\mbox{GeV/c}$,
$\theta_{\mu,\pi} < 70^{\circ}$ and $p_{\pi} < 1.6\,\mbox{GeV/c}$) 
is 38\% (42\%) if the Rein-Sehgal (Alvarez-Ruso {\it et al.}) model is used. The difference between efficiency
arises from the effect of the particle identification criterion applied to differing pion kinematic
distributions in the models.  
The cross section for scattering to the restricted phase-space is
$(3.2 \pm 0.8 (stat) \leftboth{+1.3}{-1.2}{ } (sys)) \times 10^{-40}\,\mbox{cm}^{2} / \leftexp{12}{C}\,\mbox{nucleus}$ using
the Rein-Sehgal model, and
$(2.9 \pm 0.7 (stat) \leftboth{+1.1}{-1.1}{ } (sys)) \times 10^{-40}\,\mbox{cm}^{2} / \leftexp{12}{C}\,\mbox{nucleus}$ using
the model from Alvarez-Ruso {\it et al.} These should be compared to the predictions of
$5.3 \times 10^{-40}\,\mbox{cm}^{2} / \leftexp{12}{C}\,\mbox{nucleus}$ and
$4.5 \times 10^{-40}\,\mbox{cm}^{2} / \leftexp{12}{C}\,\mbox{nucleus}$ from the models by Rein-Sehgal and Alvarez-Ruso {\it et al.},
respectively. The fractional uncertainty on these estimates from each of the main sources of systematic error are
shown in Table~\ref{tab:Exp:syst}. 
There is no guidance for the uncertainty of the coherent models in the T2K neutrino 
energy regime and so we do not include a systematic uncertainty for the signal event selection efficiency in the cross section measurement.
Fig.~\ref{fig:Exp:Q2} shows the background subtracted reconstructed $Q^{2}$ distribution compared to the two models.

Total flux-averaged cross sections may be estimated by correcting these results by the 
fraction of the full phase space contained within the restricted phase space region predicted by the model. The total
flux-averaged cross section is therefore inherently dependent on the signal model. The correction required for the two models
is 1.20 for the Rein-Sehgal model and 1.17 for the Alvarez-Ruso {\it et al.} model, leading to the total flux-averaged charged
current coherent scattering cross section of 
$(3.9 \pm 1.0 (stat) \leftboth{+1.5}{-1.4}{ } (sys)) \times 10^{-40}\,\mbox{cm}^{2} / \leftexp{12}{C}\,\mbox{nucleus}$
for the Rein-Sehgal model and 
$(3.3 \pm 0.8 (stat) \leftboth{+1.3}{-1.2}{ } (sys)) \times 10^{-40}\,\mbox{cm}^{2} / \leftexp{12}{C}\,\mbox{nucleus}$ 
in the context of the Alvarez-Ruso {\it et al.} model. These should be compared to the predictions of 
$6.4 \times 10^{-40}\,\mbox{cm}^{2} / \leftexp{12}{C}\,\mbox{nucleus}$ and
$5.3 \times 10^{-40}\,\mbox{cm}^{2} / \leftexp{12}{C}\,\mbox{nucleus}$ from the Rein-Sehgal and Alvarez-Ruso {\it et al.} models,
respectively.

It should be noted that T2K oscillation analyses utilise a version of the NEUT event generator which has undergone extensive
tuning with non-T2K neutrino scattering data and then fitted to T2K near detector data~\cite{Abe:2013xua}.
This predicts a total charged current coherent scattering flux-averaged cross section of 
$6.7 \times 10^{-40}\,\mbox{cm}^{2} / \leftexp{12}{C}\,\mbox{nucleus}$, consistent with the measurement reported 
here. By contrast, the
standard untuned NEUT predicts a total charged current coherent scattering flux-averaged cross section of 
$15.3 \times 10^{-40}\,\mbox{cm}^{2} / \leftexp{12}{C}\,\mbox{nucleus}$. The discrepancy with GENIE arises from the 
differing implementations of the pion-nucleus cross section.

{\it Conclusion}\textemdash 
T2K has made the first measurement of the cross section for charged current coherent production
of a pion from carbon nuclei for neutrino energies less than 1.5~GeV. This has been presented both in the restricted final state 
phase space ($p_{\mu, \pi} > 0.18\,\mbox{GeV/c}$,
$\theta_{\mu,\pi} < 70^{\circ}$ and $p_{\pi} < 1.6\,\mbox{GeV/c}$) and the total final state phase space. This result
disagrees with 
the null results reported previously by the K2K~\cite{Hasegawa:2005td} and SciBooNE~\cite{Hiraide:2008eu} experiments. 
These measurements have been compared to the standard Rein-Sehgal model and, for the first time, 
an instance of the class of microscopic models. While T2K observes a clear excess above background
the measured flux-averaged cross sections are below those predicted by both the Rein-Sehgal and the Alvarez-Ruso {\it et al.}
models. The statistical precision is insufficient to distinguish between the models.

\begin{acknowledgments}
We thank the J-PARC staff for superb accelerator performance and the 
CERN NA61 Collaboration for providing valuable particle production data.
We acknowledge the support of MEXT, Japan; 
NSERC (Grant No. SAPPJ-2014-00031), NRC and CFI, Canada;
CEA and CNRS/IN2P3, France;
DFG, Germany; 
INFN, Italy;
National Science Centre (NCN), Poland;
RSF, RFBR and MES, Russia; 
MINECO and ERDF funds, Spain;
SNSF and SERI, Switzerland;
STFC, UK; and 
DOE, USA.
We also thank CERN for the UA1/NOMAD magnet, 
DESY for the HERA-B magnet mover system, 
NII for SINET4, 
the WestGrid and SciNet consortia in Compute Canada, 
and GridPP in the United Kingdom.
In addition, participation of individual researchers and institutions has been further 
supported by funds from ERC (FP7), H2020 Grant No. RISE-GA644294-JENNIFER, EU; 
JSPS, Japan; 
Royal Society, UK; 
and the DOE Early Career program, USA.
\end{acknowledgments}

\bibliographystyle{apsrev4-1}
\bibliography{references}

\end{document}